\documentclass{ifacconf}

\usepackage{graphicx}      
\usepackage{natbib}        

\usepackage{amsmath}
\usepackage{amssymb}
\usepackage{booktabs}
\usepackage{wasysym}
\usepackage{mathtools}

\begin{document}
\begin{frontmatter}

\title{Modeling principles for a physiology-based whole-body model of human metabolism
}

\thanks[footnoteinfo]{This project is partially funded by Innovationfund Denmark in the project ADAPT-T2D (9068-00056B). 
Peter Emil Carstensen is funded by the The Novo Nordisk Foundation Center for Biosustainability, Technical University of Denmark, NNF20CC0035580.}

\author[DTU,NOVO]{Laura Hjort Blicher} 
\author[DTU]{Peter Emil Carstensen}
\author[NOVO]{Jacob Bendsen}
\author[NOVO]{Henrik Linden}
\author[NOVO]{Bjørn Hald}
\author[NOVO]{Kim Kristensen}
\author[DTU]{John Bagterp Jørgensen}

\address[DTU]{Technical University of Denmark, Department of Applied Mathematics and Computer Science, DK-2800 Kgs. Lyngby, Denmark}
\address[NOVO]{Novo Nordisk A/S, DK-2880 Bagsværd, Denmark}

\begin{abstract}                
Physiological whole-body models are valuable tools for the development of novel drugs where understanding the system aspects is important. This  paper presents a generalized model that encapsulates the structure and flow of whole-body human physiology. The model contains vascular, interstitial, and cellular subcompartments for each organ. Scaling of volumes and blood flows is described to allow for investigation across populations or specific patient groups. The model equations and the corresponding parameters are presented along with a catalog of functions that can be used to define the organ transport model and the biochemical reaction model. A simple example illustrates the procedure.
\end{abstract}


\end{frontmatter}
\section{Introduction}

The development of mathematical whole-body models for in-silico modelling is an important step forward in drug discovery and development in the pharmaceutical industry \citep{Moore:DrugDevelopment:2019}. These models are useful for understanding drug behavior and human metabolic processes and to inform decisions in drug discovery.

The development of a mathematical model for whole-body blood flow is driven by the need to accurately simulate the interactions between drugs, metabolism, and body functions. Traditional models are often limited by 1) their reliance on commercial platforms that are hard to automate using script languages; 2) inconsistent data sources or animal study extrapolations; and 3) their  focus on single organs without consideration of the body's integrated physiology. Automation, the ability to consistently fit parameters to data, and an integrated physiological perspective are important in understanding metabolic and drug dynamics in humans.

At the core of this effort is the integration of whole-body modelling with physiological-based pharmacokinetics (PBPK) methodologies. PBPK methodologies offers a detailed view of drug distribution and metabolism by connecting blood flow to each organ and tissue \citep{Niederalt:GenericWholeBody:2018}. This integration is key to properly represent the body's anatomical and physiological attributes, as well as the properties of both naturally occurring substances and drugs. 
 
The capacity of whole-body mathematical models to capture human metabolism has been demonstrated by \cite{Sorensen:EarlyWholeBody:1985}, \cite{Kim:MultiScale:2007}, \cite{Krauss:etal:PBPKFBA:2012}, \cite{Sluka:Liver:2016},  and \cite{Thiele:PersonalizedWholeBody:2020}. The methodology of building PBPK models is well-established and has been described extensively in the literature \citep{Willmann:WholeBody:2007, Jones:etal:basicPBPK:2013}. Nevertheless challenges remain. 
One challenge in modeling is the need to draw from various sources to establish a comprehensive framework that provides all necessary parameter values. Another challenge is to provide a transparent model and a clear understanding of the underlying equations and assumptions. Hence, the adaptability of PBPK models to diverse research needs, including those in quantitative systems pharmacology (QSP), is constrained. This paper aims to address these challenges by providing a general mathematical modeling framework for human whole-body blood flow \citep{Willmann:etal:2003, Willmann:WholeBody:2007}.  The whole-body model incorporates organ volumes and organ-specific blood flow rates obtained from the ICRP report \citep{Valentin:ICRP:2002} and PK-Sim (https://www.open-systems-pharmacology.org). It also provides allometric scaling.

We present a model structure for whole-body human physiology and integrate it with vascular, interstitial, and cellular metabolic reactions. Additionally, we offer a catalog of reaction elements. The paper is structured as follows. Section \ref{sec:WholeBodyFlowModel} introduces the whole-body blood flow model that include gender- and organ-specific volumes and flow rates. Section \ref{sec:Reactions_and_transport} details the organ compartment model by providing a catalog of reaction and transport functions. Section \ref{sec:ScalingOrganFlowVolumes} outlines the scaling functions used.
Section \ref{sec:SimpleExample} presents an application example. Section \ref{sec:Discussion} discusses the methodology. Section \ref{sec:Conclusion} summarizes the key findings.

\section{Whole-body Blood Flow Model}
\label{sec:WholeBodyFlowModel}
\begin{figure}
    \centering
    \includegraphics[scale = 0.9]{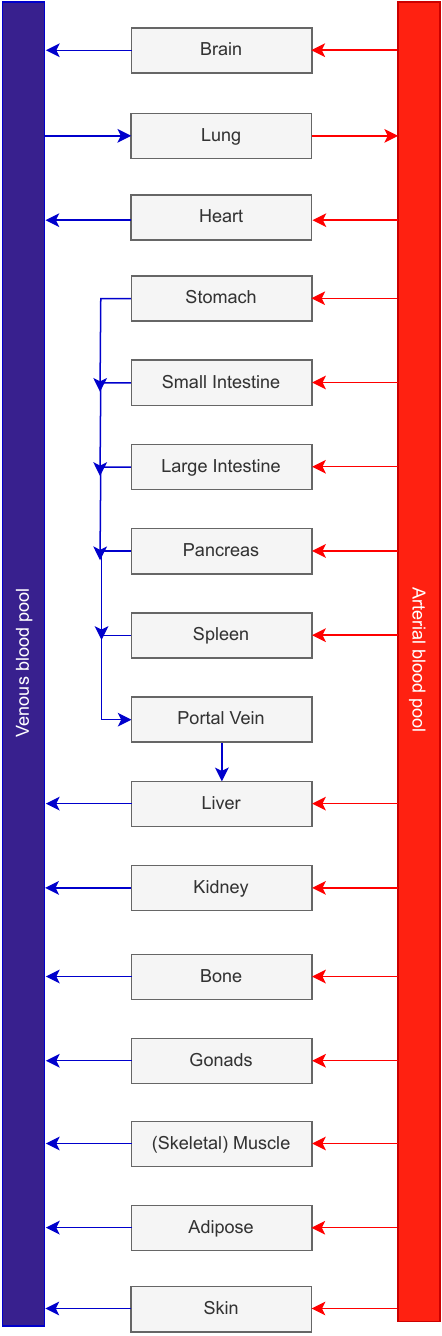}
    \caption{Schematic representation of the whole-body flow model. Arrows depict the directional movement of substance-bearing blood through the different organs, with the arterial flow shown on the right side in red and the venous return on the left side in blue. Each organ is represented as a compartment with a specific volume. The blood exiting the gastrointestinal organs and the spleen mixes in the portal vein, which then directs the flow into the liver.}
    \label{fig:whole_body_model}
\end{figure}

The human body is compartmentalized into discrete units that represent major organs or tissues. Figure \ref{fig:whole_body_model} outlines the whole-body flow diagram. Each compartment is described by a mass mass balance in the form
\begin{equation}
\label{eq:GeneralOrganMassBalanceEquation}
V \dot{C} = C_{in} Q_{in} - C Q_{out} + f
\end{equation}
$V$ is the volume, $C$ is the compartment's concentration, $C_{in}$ and $Q_{in}$ are the concentration and inflow rate, $Q_{out}$ is the outflow rate, and $f=f_R + f_T + f_S$ represents the internal metabolic rate, i.e reaction, $f_R = R V$, transport, $f_T$, and a source-sink, $f_S$, terms.
The compartments are interconnected by arterial and venous blood pools, that constitute the systemic circulation of substances throughout the body. Blood flows from arteries to veins except for the lung compartment, where the direction of flow is from venes to arteries. The lung compartment constitute the heart as a pump (but not an organ) and the pulmonary circulation for gas exchange. Accordingly, only the lung constitute the pulmonary circulation, whereas the heart compartment represents the myocardial tissue and its associated capillaries. 



The mass balance \eqref{eq:GeneralOrganMassBalanceEquation} is based on constant organ volume and assumes that the outflow from each compartment equals the sum of its inflows. This is exemplified by the portal vein compartment in Figure \ref{fig:whole_body_model}, which receives blood from the gastrointestinal organs and spleen before directing it to the liver. 



\subsection{Organ specific volumes and flow rates}
Table \ref{tab:O_vol_frac} presents the organ volumes (V) and blood flow rates (Q) for a standard ICRP European Caucasian male (73 kg, 176 cm, BMI = 23.6 kg/m$^2$) and female (60 kg, 163 cm, BMI = 22.6 kg/m$^2$). The values in Table \ref{tab:O_vol_frac} are obtained from the ICRP reference report \citep{Valentin:ICRP:2002} and PK-Sim (https://www.open-systems-pharmacology.org). The blood flow rates for the arterial and venous blood, as well as the portal vein, are not separately listed as they are derived from the cumulative inflow rates. Table \ref{tab:O_vol_frac} additionally shows the fractional organisation of organ volumes and allometric scaling values.

\begin{table*}[]
\centering
\caption{
Organ volumes (V) and blood flow rates (Q) of an adult ICRP European Caucasian. Organ composition fractions are divided into vascular space (v), interstitial space (i), and intracellular space (c). Organ composition fractions are obtained from PK-sim Version 11.2 Build 142, while cardiac output fractions ($\kappa_O$) are calculated as described in \cite{Willmann:WholeBody:2007}. The allometric scaling factor $\bar{\beta}$ is from PK-Sim documentation, and $\bar{\alpha}$ and $\bar{\gamma}$ are from \cite{Willmann:WholeBody:2007}.}
\begin{tabular}{llllllllllllll} \toprule
                &    & \multicolumn{2}{c}{{V (L)}} & \multicolumn{2}{c}{Q (L/min)} &  \multicolumn{3}{c}{{$V_{Ok}/V_{O}, \, k\in\{v,i,c\}$}}   & \multicolumn{2}{c}{{$\kappa_O = Q_O/Q_{LU}$}}  & $\bar{\alpha}$ & $\bar{\beta}$ & $\bar{\gamma}$ \\ 
{Organ}         & O  & Male   & Female  & Male    & Female  & v       & i      & c      & Male    & Female                    &      &      &  \\ \hline
Arterial blood  & A  & 0.419  & 0.320   & -       & -       & 1       & 0      & 0      &  -      & -                         & 0.00 & 0.75 & 0.00 \\
Venous blood    & V  & 0.964  & 0.737   & -       & -       & 1       & 0      & 0      &  -      & -                         & 0.00 & 0.75 & 0.00 \\ 
Portal vein     & PV & 1.037  & 0.793   & -       & -       & 1       & 0      & 0      & -       & -                         & 0.00 & 0.75 & 0.00 \\ \hline
Adipose         & AD & 14.65  & 19.42   & 0.320   & 0.503   & 0.018   & 0.160  & 0.822  & 0.053   & 0.092                     & 0.00 & 0.00 & 0.00 \\
Brain           & B  & 1.509  & 1.355   & 0.780   & 0.706   & 0.039   & 0.004  & 0.957  & 0.128   & 0.129                     & 0.00 & 0.00 & 0.00 \\
Bone            & BO & 11.82  & 9.122   & 0.325   & 0.295   & 0.034   & 0.100  & 0.866  & 0.053   & 0.054                     & 0.00 & 2.00 & 0.00 \\
Gonads          & G  & 0.040  & 0.013   & 0.003   & 0.001   & 0.055   & 0.069  & 0.876  & $5\cdot 10^{-4}$ & $2\cdot 10^{-4}$ & 0.00 & 0.75 & 0.00 \\
Heart           & H  & 0.417  & 0.329   & 0.260   & 0.295   & 0.140   & 0.100  & 0.760  & 0.042   & 0.054                     & 0.00 & 0.75 & 0.00 \\
Kidney          & K  & 0.438  & 0.404   & 1.327   & 1.122   & 0.230   & 0.200  & 0.570  & 0.218   & 0.205                     & 0.00 & 0.75 & 0.00 \\
Liver           & L  & 2.377  & 1.918   & 0.426   & 0.386   & 0.170   & 0.163  & 0.667  & 0.069   & 0.070                     & 0.00 & 0.75 & 0.00 \\
Large intestine & LI & 0.413  & 0.417   & 0.260   & 0.295   & 0.024   & 0.094  & 0.882  & 0.043   & 0.054                     & 0.00 & 0.75 & 0.00 \\
Lung            & LU & 1.215  & 0.946   & 6.088   & 5.482   & 0.580   & 0.188  & 0.232  & 1       & 1                         & 0.00 & 0.75 & 0.75 \\
Muscle          & M  & 32.65  & 20.29   & 1.116   & 0.666   & 0.025   & 0.160  & 0.815  & 1.832   & 0.121                     & 0.00 & 2.00 & 0.00 \\
Pancreas        & P  & 0.190  & 0.170   & 0.065   & 0.059   & 0.200   & 0.120  & 0.680  & 0.011   & 0.012                     & 0.00 & 0.75 & 0.00 \\
Small intestine & SI & 0.725  & 0.695   & 0.650   & 0.649   & 0.024   & 0.094  & 0.882  & 0.107   & 0.118                     & 0.00 & 0.75 & 0.00 \\
Skin            & SK & 3.763  & 2.724   & 0.325   & 0.296   & 0.046   & 0.302  & 0.652  & 0.053   & 0.054                     & 0.50 & 0.75 & 0.00 \\
Spleen          & SP & 0.207  & 0.186   & 0.166   & 0.150   & 0.330   & 0.150  & 0.520  & 0.027   & 0.027                     & 0.00 & 0.75 & 0.00 \\
Stomach         & ST & 0.169  & 0.163   & 0.065   & 0.059   & 0.032   & 0.100  & 0.868  & 0.011   & 0.011                     & 0.00 & 0.75 & 0.00 \\ 
\bottomrule
\end{tabular}%
\label{tab:O_vol_frac}
\end{table*}

\subsection{Organ Compartments}  

\begin{figure}
    \centering
    \includegraphics[width = \columnwidth]{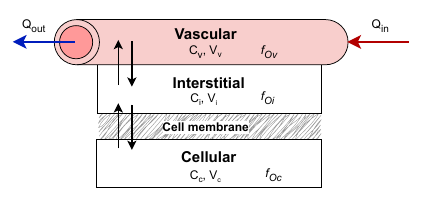}
    \caption{Schematic representation of an organ with three subcompartments; a vascular, interstitial, and intracellular compartment. The transport between subcompartments is defined by $T_{O}$ and the blood flow between organs is defined by $q=C\cdot Q$. 
    }
    \label{fig:Organ_model}
\end{figure}

Organ compartments are sub-divided into three sub-compartments that describe the vascular, interstitial, and intracellular space of the organ.
Figure \ref{fig:Organ_model} outlines the sub-divisions of the organ compartment. Consider the set of all organs, 
$\mathcal{O} = $ $\{$A, V, PV, AD, B, BO, G, H, K, L, LI, LU, M, P, SI, SK, SP, ST$\}$. Then the mass balances for the vascular part can be represented as 

\begin{subequations}
\begin{alignat}{3}
    \label{eq:DC_v}
    V_{Ov} \dot{C}_{Ov} &= q_{O} - C_{Ov} Q_{O,out} + f_{Ov} 
\qquad && \forall O \in \mathcal{O} \\
    q_{O} &= \sum_{\bar O \in \mathcal{I}_O} C_{\bar Ov} Q_{\bar O} \qquad && \forall O \in \mathcal{O} \\
    Q_{O,out} &= \sum_{\bar O \in \mathcal{I}_O} Q_{\bar O} && \forall O \in \mathcal{O} 
\end{alignat}
\end{subequations}
The mass balances for the interstitial organ spaces and the cellular organ spaces can be represented as
\begin{subequations}
\begin{alignat}{2}
    \label{eq:DC_i}
    V_{Oi} \dot{C}_{Oi} &= f_{Oi} \qquad && \forall O \in \mathcal{O} \\
    \label{eq:DC_c}
    V_{Oc} \dot{C}_{Oc} &= f_{Oc} && \forall O \in \mathcal{O}
\end{alignat}
\end{subequations}
$V_{Ov}$ is the vascular volumes, $V_{Oi}$ is the interstitial volumes and $V_{Oc}$ is the cellular volumes defined by the fractions outlined in Table \ref{tab:O_vol_frac}. $f_{Ov}$, $f_{Oi}$ and $f_{Oc}$ are the vascular, interstitial and cellular molar transport and production (reaction) rates. $\mathcal{I}_O$ is the set of inflows to organ $O$. The subscript $v$ indicates the vascular space, $i$ the interstitial space, and $c$ the intracellular space. Appendix \ref{app:ODEsFlowStructure} outlines the vascular compartments for each organ.


The choice of sub-compartments to be modeled depends on the physiological context and the modeling task. For instance, arteries and veins, primarily involved in the transport of blood, may be represented by only the vascular compartment. Conversely, when modeling the immune response, the interstitial space becomes crucial, as it serves as a conduit for immune cell trafficking. Similarly, the intracellular space is essential for depicting intracellular processes such as energy metabolism, or sub-cellular drug targeting. 

\subsection{Transport and reaction rates}
The molar transport and reaction rates in the vascular, interstitial, and cellular spaces are $f_{Ov}$, $f_{Oi}$, and $f_{Oc}$. They can be computed as the sum of transport and reaction rates for each organ ($\forall O \in \mathcal{O}$),
\begin{subequations}
\begin{alignat}{3}
f_{Ov} &= -T_{Ovi} V_{Ov} + T_{Oiv} V_{Oi} + R_{Ov} V_{Ov} + s_{Ov} \\
\begin{split}
f_{Oi} &= T_{Ovi} V_{Ov} - T_{Oiv} V_{Oi} \\ & \quad - T_{Oic} V_{Oi} + T_{Oci} V_{Oc} + R_{Oi} V_{Oi}  + s_{Oi}
\end{split} \\
f_{Oc} &= T_{Oic} V_{Oi} - T_{Oci} V_{Oc} + R_{Oc} V_{Oc}  + s_{Oc}
\end{alignat}
\end{subequations}
The transport across membranes in each organ may be defined by equations for rates from vascular to interstitial, $T_{Ovi}$, interstitial to vascular, $T_{Oiv}$, interstitial to cellular, $T_{Oic}$, and cellular to interstitial, $T_{Oci}$, spaces. The transport functions  
\begin{subequations}
\label{eq:OrganTransportRateModelGeneral}
\begin{alignat}{3}
    T_{Ovi} &= T_{Ovi}(C_{Ov}) \qquad && \forall O \in \mathcal{O} \\
    T_{Oiv} &= T_{Oiv}(C_{Oi})\qquad && \forall O \in \mathcal{O} \\
    T_{Oic} &= T_{Oic}(C_{Oi}) \qquad && \forall O \in \mathcal{O} \\
    T_{Oci} &= T_{Oci}(C_{Oc})\qquad && \forall O \in \mathcal{O} 
\end{alignat}
\end{subequations}
could be diffusion, $T = k C$, or functions related to active transport, e.g. $T = \mu C/(K+C)$.    
Equation \eqref{eq:OrganTransportRateModelGeneral} is denoted the organ transport model.

The production rates in each subcompartment 
\begin{subequations}
\label{eq:ChemicalModel}
\begin{alignat}{3}
R_{Ov} &= \nu_{Ov}' r_{Ov}(C_{Ov}) \qquad && \forall O \in \mathcal{O} \\
R_{Oi} &= \nu_{Oi}' r_{Oi}(C_{Oi}) \qquad && \forall O \in \mathcal{O} \\
R_{Oc} &= \nu_{Oc}' r_{Oc}(C_{Oc}) \qquad && \forall O \in \mathcal{O}
\end{alignat}
\end{subequations}
are obtained from the stoichiometry and reaction rates for each compartment ($v,i,c$) in all organs $\mathcal{O}$. We call \eqref{eq:ChemicalModel} the organ biochemical model.

The terms $s_{Ov}$, $s_{Oi}$, and $s_{Oc}$ denote source-sink functions in the vascular, interstitial, and cellular compartments. These terms may used to model the removal of a substance, e.g. drug clearance in the kidneys, and uptake of a substance as in uptake of macro-nutrients (carbohydrates, proteins, lipids) in the small intestine (SI) compartment from the gastro-intestinal tract \citep{Ritschel:etal:MealModels:2023}. 

The main modeling task is the definition of the organ transport rate model and the organ biochemical model. This corresponds to specification of the transport rate function, $T(C)$, the stoichiometric matrix, $\nu$, and the corresponding reaction rate function, $r(C)$, for all compartments ($v,i,c$) in all organs for $O \in \mathcal{O}$ \citep{Carstensen:Bendsen:etal:FOSBE:2022}.

\section{Reaction and transport functions}
\label{sec:Reactions_and_transport}
We illustrate the principles for construction of the biochemical model \eqref{eq:ChemicalModel}. Similar principles can be applied in the construction of the transport model \eqref{eq:OrganTransportRateModelGeneral}. Consider a set of reactions specified by the reaction stoichiometry and kinetics    
\begin{equation}
\sum_{j \in \mathcal{C}} \alpha_{ij} S_j \rightarrow \sum_{j \in \mathcal{C}} \beta_{ij} S_j, \qquad r_i = r_i(C), \qquad \forall i \in \mathcal{R}
\end{equation}
The stoichiometric coefficient for component $j$ in reaction $i$ is $\nu_{ij} = \beta_{ij}-\alpha_{ij}$ $\forall i \in \mathcal{R}$ and $\forall j \in \mathcal{C}$. The stoichiometric matrix is $\nu = [\nu_{ij}]_{i\in \mathcal{R},j\in \mathcal{C}} = \beta - \alpha$. The production rate vector is $R = [R_j]_{j \in \mathcal{C}}$ and the reaction rate vector is $r = [r_i]_{i \in \mathcal{R}}$ with $r=r(C)$. Then
\begin{equation}
R = \nu' r, \quad r = r(C)
\end{equation}
The choice of the rate functions, $r(C)$, depends on the biological mechanisms in the reaction being modelled. 
The kinetic function for a reaction is typically in the form
\begin{equation}
    r_i(C) = k_i \prod_{j \in \mathcal{C}_i} {\xi_{ij}(C_j)} \quad \forall i \in \mathcal{R}
\end{equation}
where $k_i$ is a constant and $\xi$ are different basis functions. The basis functions includes the \textit{power law} 
\begin{equation}
    \xi_{ij} = \xi_{ij}(C_j) = C_j^{\alpha_{ij}}
\end{equation}
The basis functions may also be kinetic models from biochemistry, pharmacology, and microbiology. \\
\textit{Michaelis-Menten (Monod)}
\begin{subequations}
\begin{equation}
    \xi_{ij} = \xi_{ij}(C_{j}) = \mu_{ij} \frac{C_{j}}{K_{50,ij}+C_{j}}
\end{equation}
\textit{Hill}
\begin{equation}
    \xi_{ij} = \xi_{ij}(C_j) = \mu_{ij} \frac{ C_{j}^{n_{ij}} } { K_{50,ij}^{n_{ij}} + C_{j}^{n_{ij}}}
\end{equation}
\textit{Monod-Haldane (Inhibition)}
\begin{equation}
    \xi_{ij} = \xi_{ij}(C_{j}) = \mu_{ij} \frac{C_{j}}{K_{50,ij} + C_{j} + C_{j}^2/K_{I,ij}}
\end{equation}
\end{subequations}
\textit{Sigmoid}\footnote{See \tt{https://en.wikipedia.org/wiki/Sigmoid\_function}} basis functions such as $\xi = \xi(x) = \tanh(x)$ (can be regarded a smooth approximation to 
$\xi(x) = \text{sgn}(x)$) scaled to specific starting and activation concentrations
\begin{subequations}
\begin{align}
    &\xi_{ij} = \xi_{ij}(C_j) = A_{ij} \tanh(c_{ij}(C_j-d_{ij})) + B_{ij}\\
    & \qquad A_{ij} = \frac{a_{ij}-b_{ij}}{2}, \quad B_{ij} = \frac{a_{ij}+b_{ij}}{2} 
\end{align}
\end{subequations}
are frequently used. $a$ is the limit for $C \rightarrow -\infty$, $b$ is the limit for $C \rightarrow \infty$, $c$ is the steepness and $d$ is the location of the half maximum. Furthermore, the soft-max\footnote{\tt{https://en.wikipedia.org/wiki/Softmax\_function}} and soft-plus\footnote{\tt{https://en.wikipedia.org/wiki/Rectifier\_(neural\_networks)}} functions are often used. As an example, the soft-plus functions $\xi(C) = 0.5 \left( C + \sqrt{\beta^2 + C^2} \right)$ and $\xi(C) = \beta^2 \ln \left( 1+\exp(C/\beta^2) \right)$ are smooth approximations to $\xi(C) = C^+ = \max\{0, C \}$ (the approximations are exact for $\beta \rightarrow 0$). Sigmoid, soft-max, and soft-plus functions are also used as basis functions in machine learning.

\section{Scaling of organ volumes and flows}
\label{sec:ScalingOrganFlowVolumes}
Scaling of organ volumes and blood flow is necessary to adjust the model to individual body characteristics like weight and height. Allometric scaling is a widely accepted method for adjusting organs sizes and physiological flows to adjust for population variations \citep{Willmann:WholeBody:2007}. For most organs, the scaling of organ volume is based on the following allometric equation
\begin{gather}
    \frac{V_O}{\bar V_O} =  \left( \frac{BW}{\bar{BW}} \right)^{\alpha_O} \left( \frac{h}{\bar h} \right)^{\beta_O}, \\
    \quad \alpha_{O} = \bar{\alpha}_{O}, \quad \beta_O = \bar{\beta}_O,  \quad O \in \mathcal{O}\setminus \{ AD \} \nonumber
\end{gather}
$V_O$ is the scaled volume for an individual of a specific body weight, $BW$, and height, $h$. The mean values are denoted with a bar \citep{Valentin:ICRP:2002}. Table \ref{tab:O_vol_frac} provides the parameters.

The adipose tissue compartment is used as a 'buffer' to ensure that the desired body weight is reached. Assume that the density is $\rho =$ 1 g/cm$^3$ = 1 kg/L. Then the adipose volume is scaled by the formula
\begin{equation}
    \label{eq:V_AD}
    V_{AD} = V_{tot} - \sum_{O \in \mathcal{O}\setminus \{ AD \}} V_O,  \quad V_{tot} = BW/\rho
\end{equation}
Organ blood flow rates are allometrically scaled with respect to height as \citep{Willmann:WholeBody:2007}   
\begin{subequations}
\begin{alignat}{3}
   \frac{Q_{LU}}{\bar{Q}_{LU}} &= \left( \frac{h}{\bar h} \right)^{\gamma_{LU}}, \quad && \gamma_{LU} = 0.75 \\
    {Q}_{O} &= Q_{LU}  \kappa_O,  \quad && O \in \mathcal{O}
\end{alignat}
\end{subequations}
$Q_{LU}$ represents the scaled total cardiac output. Table \ref{tab:O_vol_frac} provides the parameters.
\section{Example Application}
\label{sec:SimpleExample}
\begin{figure*}[h!]
\centering
\includegraphics[width = 0.89\textwidth]{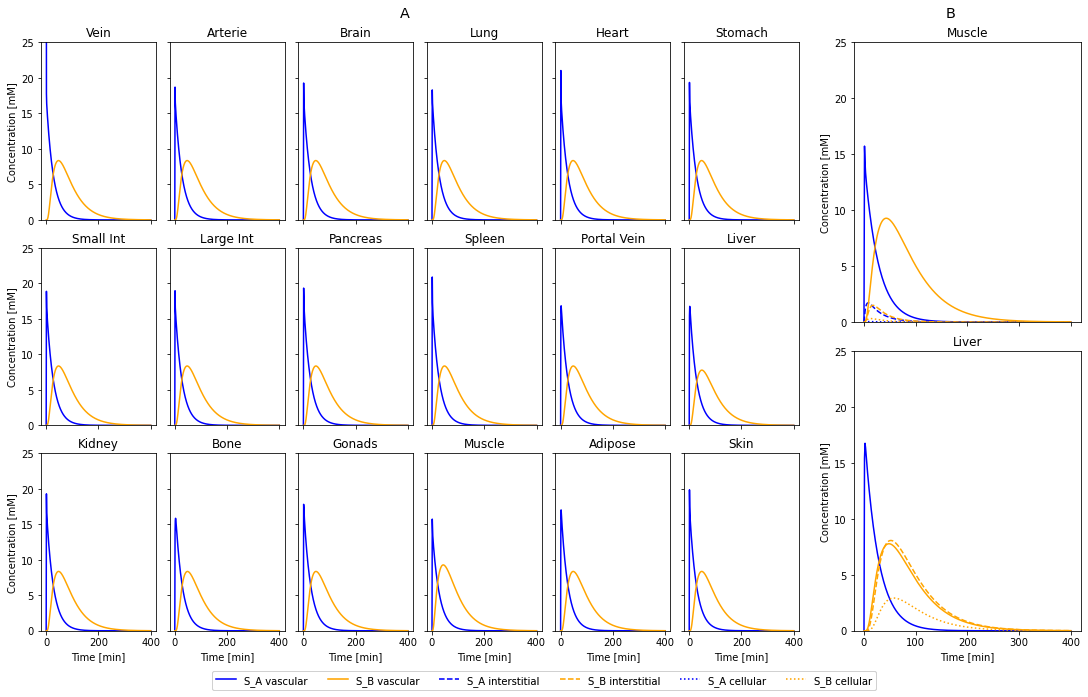}
\caption{A shows the distribution of $S_A$ and $S_B$ across the vascular compartments in the whole body over time. Blue line is $S_A$ and orange line is $S_B$. B shows the distribution of $S_A$ and $S_B$ in the vascular, interstitial and cellular space. 
}
\label{fig:Example_flow}
\end{figure*}

By a simple example, we demonstrate the distribution of two substrates, $A$ $(S_A)$ and $B$ $(S_B)$, in a standard male. $A$ enters the blood stream through a bolus injection in the vein such that $C_{Vv,{A}}(0) = 100$ mM. It disperses to organ. In the muscle, it moves from vascular to interstitial to cellular spaces and metabolizes to $B$. $B$ circulates from muscle to blood, distributes, and is finally cleared in the cellular space of the liver. Table \ref{tab:example_transport} defines the organ transport model, $T(C)$. Table  \ref{tab:example_stoichio_table} defines the organ biochemical model, $\nu$ and $r(C)$. The corresponding stoichiometric matrices are $\nu_{Mc} = \begin{bmatrix} -1 & 1 \end{bmatrix}$ and $\nu_{Lc} = \begin{bmatrix} 0 & -1 \end{bmatrix}$, respectively. Figure \ref{fig:Example_flow} shows the evolution of $A$ ($S_A$) and $B$ ($S_B$) in all organs over time.

\begin{table}
\centering
 \caption{The organ transport model, $T(C)$, in the example model. }
 \label{tab:example_transport}
  \begin{tabular}{clll}
   \toprule
    Species  & Organ   & Kinetic & Parameters \\[0.5ex]
    \midrule
    $A$           & M & $T_{M,vi} = k_1 C_{Mv,A}$  & $k_1 = 0.3$ min$^{-1}$ \\
    $A$           & M & $T_{M,ic} = k_2 C_{Mi, A}$  & $k_2 = 0.3$ min$^{-1}$ \\
     \midrule
    $B$           & M & $T_{M,ci} = k_3 C_{Mc, B}$  & $k_3 = 0.3$ min$^{-1}$  \\
    $B$           & M & $T_{M,iv} = k_4 C_{Mi, B}$  & $k_4 = 0.3$ min$^{-1}$ \\
    \midrule
    $B$           & L & $T_{L,vi} = k_5 C_{Lv, B}$  & $k_5 = 0.3$ min$^{-1}$  \\
    $B$           & L & $T_{L,ic} = k_6 C_{Li, B}$  & $k_6 = 0.3$ min$^{-1}$  \\
   \bottomrule
  \end{tabular}
\end{table}

\begin{table}
\centering
 \caption{The organ biochemical model, stoichiometry ($\nu$) and reaction kinetics, $r(C)$, in the example model. $\mu_1 = 2$ mM/min, $K_{50} = 0.8$ mM and  $\mu_2 = 0.2$ min$^{-1}$.}
 \label{tab:example_stoichio_table}
  \begin{tabular}{cccll}
   \toprule
    \#  & Organ & Space & Stoichiometry   & Kinetic  \\
     [0.5ex]
   \midrule
    1      &M & c      & $A \rightarrow B $ & $r_1$ = $\mu_1 \frac{C_{A}}{K_{50} +C_{A}}$  \\
    \midrule
    2      &L & c     & $B \rightarrow \varnothing$ & $r_2$ = $\mu_2 C_{B}$ \\
   \bottomrule
  \end{tabular}
\end{table}

\section{Discussion}
\label{sec:Discussion}
In this paper, we present a whole-body model with 18 organ compartments, combined with transport and biochemical models, to provide a comprehensive representation of human physiology and organ-specific metabolism. The framework forms the basis for detailed metabolic simulations, incorporating a suite of smooth rate equations for computational stability and efficiency. The model's adaptable design permits the application of different methodologies, such as flux balance analysis \citep{Thiele:PersonalizedWholeBody:2020, Toroghi:etal:WBM:2016} and standard pharmacokinetic equations \citep{Jusko:PKPDmodels:2013}, to suit a range of modeling purposes and research objectives to aid drug discovery.

Accurate organ volumes and blood flow rates are important for physiological models. The model in this paper uses ICRP reference data for healthy European adults. In Section \ref{sec:ScalingOrganFlowVolumes}, we present scaling functions for body weight and height to be able to model an adult population \citep{Willmann:WholeBody:2007}. This scaling enables modeling of conditions like obesity that affect drug kinetics \citep{Gouju:Obesity:2023}. However, the scaling functions are not precise as the organ sizes in a population can vary significantly. The scaling functions used in our model assume that weight changes are primarily due to gain in adipose tissue. This may not be accurate for individuals with significant muscle mass. These variations in body composition affect drug distribution and clearance predictions. The model's reliance on population averages also overlooks individual-specific factors. This underscores the need for cautious application and further refinement to accommodate diverse body compositions.

In physiological whole-body modeling, organs are typically divided into three sub-compartments for detailed flow and interaction modeling \citep{Thiele:PersonalizedWholeBody:2020, Sorensen:EarlyWholeBody:1985}. Some models use a four-compartment approach that separates the plasma and the red blood cells within the vascular compartment \citep{Willmann:WholeBody:2007}. The choice between three or four compartments depends on the desired balance between complexity and detail. While the three-compartment model is often adequate for many studies, the four-compartment model is preferable for analyzing specific blood component interactions. However, increased compartmentalization can complicate computations and parameter estimation, requiring a balance between resolution and feasibility.

\section{Conclusion}
\label{sec:Conclusion}
We have presented a whole-body modeling structure that includes organ-specific flow and volume parameters. Each organ contains a vascular, an interstitial, and  a cellular space. The main modeling task for a specific case is to define the organ transport and the organ biochemical reactions.  We believe, that the model structure presented will serve as a valuable tool for exploring integrated systems and cellular aspects in human physiology and aid in drug discovery.

\bibliography{ifacconf}

\appendix
\section{Mass balances for the vascular compartments in each organ}
\label{app:ODEsFlowStructure}
Consider the set of all organs, $\mathcal{O} = $ $\{$A, V, PV, AD, B, BO, G, H, K, L, LI, LU, M, P, SI, SK, SP, ST$\}$. Then the mass balance equations for the vascular organ space can be represented as: 
\begin{gather}
    V_{Ov} \dot{C}_{Ov} = (C_{Av} - C_{Ov}) Q_O + f_{Ov},  \\
    O \in \mathcal{O}\setminus \{ A, V, PV, L, LU\}
\end{gather}
Artery (A): 
\begin{align}
& V_{Av} \dot{C}_{Av} = (C_{LUv} - C_{Av}) Q_{LU} + f_{Av} 
\end{align}
Vein (V): 
\begin{subequations}
\begin{align}
& V_{Vv} \dot{C}_{Vv} = q_{V} - C_{Vv} Q_V + f_{Vv} \\
\begin{split}
    &q_V = C_{Bv} Q_B + C_{Hv} Q_{H} + C_{Lv} Q_{L,out}  \\
    & \qquad \quad + C_{Kv} Q_K + C_{Gv} Q_{G} + C_{BOv} Q_{BO} \\
    & \qquad \quad + C_{Mv} Q_M + C_{Fv} Q_F + C_{SKv} Q_{SK} 
\end{split}
\\
\begin{split}
    &Q_V = Q_B + Q_H + Q_{L,out} + Q_K + Q_{G}   
    \\& \qquad \quad +Q_{BO}+ Q_M + Q_F + Q_{SK} 
\end{split}
\end{align}
\end{subequations}
Portal vein (PV):
\begin{subequations}
\begin{align}
& V_{PVv} \dot{C}_{PVv} = q_{PV} - C_{PVv} Q_{PV} + f_{PVv} \\
\begin{split}
 & q_{PV} = C_{STv} Q_{ST} + C_{SIv} Q_{SI} + C_{LIv} Q_{LI} 
\\ & \qquad \quad + C_{Pv} Q_{P} + C_{SPv} Q_{SP} 
\end{split}
\\
&Q_{PV} = Q_{ST} + Q_{SI} + Q_{LI} + Q_{P} + Q_{SP}
\end{align}
\end{subequations}
Liver (L):
\begin{subequations}
\begin{align}
    &V_{Lv} \dot{C}_{Lv} = q_L - C_{Lv} Q_{L,out} + f_{Lv}
\\
&q_L = C_{PVv} Q_{PV} + C_{Av} Q_L 
\\
& Q_{L,out} = Q_{PV} + Q_L 
\end{align}
\end{subequations}
Lung (LU): 
\begin{equation}
V_{LUv} \dot{C}_{LUv} = (C_{Vv} - C_{LUv}) Q_{LU} + f_{LUv}
\end{equation}

\end{document}